\documentclass[12pt,letterpaper]{article}
\usepackage{amsmath,amssymb,pgf,pgfarrows,pgfnodes,float,appendix, hyperref}
\usepackage{graphicx}
\usepackage{subfigure}
\usepackage[margin=0.9in]{geometry}

\title{{\rm\footnotesize \qquad \qquad \qquad \qquad \qquad \ \qquad \qquad \qquad \ \ \ \ \ \                  RUNHETC-2018-31, UTTG-24-18 }\vskip.5in    Why The Cosmological Constant is a Boundary Condition}
\author{Tom Banks\\
Department of Physics and NHETC\\
Rutgers University, Piscataway, NJ 08854\\
E-mail: \href{mailto:tibanks@ucsc.edu}{tibanks@ucsc.edu}
\\
\\
Willy Fischler\\
Department of Physics and Texas Cosmology Center\\
University of Texas, Austin, TX 78712\\
E-mail: \href{mailto:fischler@physics.utexas.edu}{fischler@physics.utexas.edu}}
\date{}
\begin{document}
\maketitle

\begin{abstract}
We review arguments that the cosmological constant (c.c.) should not be thought of as a local contribution to the energy density, but rather as an infrared boundary condition specifying particular models of quantum gravity. \end{abstract}

\section{Introduction}

The Cosmological Constant Problem has haunted high energy physics for decades \cite{weinbergcdl} \footnote{One of the authors (TB) first heard about this problem in lectures by K. Johnson at the 1976 Scottish Summer School.}.  The traditional way to think about this problem is in bulk effective field theory.  Every such model has a stress tensor $T_{\mu\nu}$, and a unique ground state, and general covariance implies that if one calculates the expectation value of the stress tensor in the ground state, it will be proportional to $g_{\mu\nu} (x)$, where $g_{\mu\nu}$ is the space-time metric on which the field theory lives.  If we now imagine that $g_{\mu\nu}$ 
is itself a dynamical quantum field, this leads to a change in the c.c. term in the gravitational Lagrangian.  Power counting arguments show that if we calculate this term by standard field theory rules, the value of the "correction" to the c.c. is of order $M_c^d$, where $M_c$ is of order the UV cutoff.  The apparent value of the c.c. in the real world could only be compatible with this estimate if $M_c \sim 10^{-3}$ eV. 
The only known symmetry that can avoid this disaster is supersymmetry.  Supersymmetry is consistent only with non-positive c.c. and the c.c. appears in the SUSY algebra and cannot be renormalized.  The experimental lower bound on SUSY violation implies that this cannot be the explanation for the value of the c.c. in the real world.

We have long advocated an alternative conceptual view of the c.c. as an infrared (IR) boundary condition that is part of the intrinsic definition of models of quantum gravity \footnote{For an approach which attempts to determine the numerical value of the cosmological constant as an inital condition at cosmogenesis, see \cite{tp3}.}. The purpose of the present note is to review the arguments that this is the proper way to define the c.c., and that the field theoretic calculations stretch the quantum field theory approximation to quantum gravity beyond its range of validity.  At the heart of our understanding is the question of how the quantum field theory approach to "quantizing" Einstein's classical theory of gravitation emerges from an underlying and more fundamental theory.  We believe that the strongest clue we have to the nature of its emergence is Jacobson's\cite{ted} 1995 derivation of the Einstein equations, {\it without the c.c.}, from the hydrodynamics of the generalized Bekenstein-Hawking area law relating quantum entropy to areas of the holographic screens of causal diamonds.
This derivation of General Relativity from hydrodynamics leads one to recall that hydrodynamics should only be quantized in situations of relatively low entropy and energy, where the only relevant states are quantized excitations of hydrodynamic fields.  In ordinary condensed matter physics, the high entropy part of the spectrum is still described by classical hydrodynamics, even though the quantum excitations that give rise to the entropy, have nothing to do with quantized hydrodynamics.  We will see that the c.c. puzzle is intimately related to this hydrodynamic view of Einstein's equations within models of quantum gravity.

The second part of our paper will be devoted to the evidence from AdS/CFT that the c.c. is a parameter related to I.R. boundary conditions, which, via the well known UV/IR duality, is the UV definition of the boundary CFT.  We have been somewhat surprised that these well known and completely agreed upon facts about AdS/CFT have not changed the community's view of the bulk c.c. problem in field theory.

Finally, we will point out how certain well known facts about horizons find no explanation within conventional quantum field theory treatments.  These lead to a simple picture of how quantum field theory emerges from a theory of quantum gravity based on the entropy formula, in which localized objects are constrained states of boundary degrees of freedom, and the interpretation of the c.c. as a boundary condition is manifest.

\section{The Jacobson-Einstein Equation of State}

Let us briefly sketch Jacobson's seminal derivation of the gravitational field equations from the entropy formula\footnote{We will be slightly anachronistic, and use Bousso's description of the BHJFSB\cite{BHJFSB} entropy bound in terms of holographic screens to describe Jacobson's work.} for causal diamonds.  A causal diamond in space-time is the region bounded by the intersection of the backward lightcone of a point $P$ and the forward light cone of some point in the causal past of $P$.  It has two characteristic scales associated with it, the proper time of the geodesic between its tips, and the area of the maximal area leaf in a null foliation of its boundary.  The Covariant Entropy Principle states that the maximal entropy associated with the diamond is given by one quarter of the area of that leaf in Planck units.

Jacobson assumes that the radii of curvature of the space-time geometry are much larger than any fundamental microscopic scale, so that we can concentrate on a macroscopic region where the geometry is approximately flat.  Now we want to contemplate small changes in the entropy concentrated near a particular point in space-time, which we take to lie on the maximal area leaf of some causal diamond.  Hydrodynamics is the local form of the equation 
\begin{equation} dE = T dS . \end{equation}  To identify $E$ and $T$, Jacobson invokes Unruh's classic work  on the temperature due to acceleration.  In order to concentrate attention just on a single point, Jacobson takes the infinite temperature limit of an Unruh trajectory that grazes the boundary of the causal diamond, with turnaround at $P$.  This has two important consequences.  An infinite temperature limit implies that we are counting all of the entropy of the Hilbert space associated with the diamond, so that Hilbert space is finite dimensional and the entropy is the logarithm of its dimension.  Secondly, the energy in the local first law of thermodynamics is the energy measured along the Unruh trajectory $k^{\mu} (P) k^{\nu} (P) T_{\mu\nu} (P)$, so $k^{\mu}$ is null.  
The change in entropy can be evaluated by Raychaudhuri's equation, and Jacobson shows that this leads to (in $4$ dimensions)
\begin{equation}  k^{\mu} (P) k^{\nu} (P) (R_{\mu\nu} - \frac{1}{2} g_{\mu\nu} R - 8\pi G_N T_{\mu\nu})  = 0 . \end{equation}  Applying this reasoning at every point and every maximally accelerated Unruh trajectory, we have derived Einstein's equations, but without the c.c. .

The four obvious lessons to be learned from this remarkable argument are 
\begin{itemize}
\item The c.c. is not a local energy density.  It does not contribute to the hydrodynamic equations \footnote{The relevance of Eq(1), in contrast to the standard Einstein's equations, for the cc problem has been emphasised by Padmanabhan in several papers; see e.g. \cite{tp1, tp2}}.
\item The entropy invoked in the Covariant Entropy Principle is the logarithm of the dimension of the Hilbert space associated with the diamond.  This was stated in one of Bousso's papers\cite{bousso} and we have advocated it in our own work, but Jacobson's argument is the definitive reason to believe it is correct.
\item Einstein's equations are hydrodynamic equations and should only be quantized in very specific low energy/low entropy circumstances.
\item Even an "empty" causal diamond in flat space has entropy proportional to its area.  Unruh's argument shows that this entropy is visible along certain trajectories in flat space\footnote{The specifics of Unruh's derivation are carried out in quantum field theory, and are related to the Type III property of quantum field theory algebras. In fact, there is a derivation of Unruh's result along the lines of Hartle and Hawking's derivation of black hole temperature\cite{hh}.  This relies only on the geometric fact that the continuation of a hyperbola to imaginary time is a circle.  Thus, Unruh's result really reflects a basic fact about geometry, and not the specific quantum system under study.}.   We'll see later that in de Sitter space it is visible even to the geodesic trajectory.
\end{itemize}

Finally let us note a hint about the eventual formulation of quantum gravity, provided by Jacobson's derivation.  One is encouraged to think about the Hamiltonian of quantum gravity as the generator of translations in proper time, along a particular time-like trajectory.

\section{AdS/CFT}

The AdS/CFT correspondence is widely and correctly viewed as the most complete non-perturbative model of quantum gravity in existence.  There are a number of important facts about this correspondence that are relevant to our discussion.  Perhaps the most salient is the relation between the c.c. and the entropy of large AdS black holes
\begin{equation} S = \pi (2M R)^{2/3} (R/L_P)^{2/3} , \end{equation} with analogous equations in other dimensions.  This connects the c.c. to the {\it high} energy behavior of the CFT spectrum and can also be viewed as the first hint that the boundary Hamiltonian is a quantum field theory.  Since the black hole radius is large in the limit $MR \rightarrow\infty$, this equation also incorporates the UV/IR correspondence. This quite explicitly disagrees with the bulk effective field theory notion that the c.c. is just the low energy vacuum energy of the theory.   The vacuum energy of the CFT vanishes, and the energies of low energy states are proportional to $1/R$ only because we've used conformal invariance to set that as the energy scale of the scale free theory.  

In quantum mechanics, the definition of a theory is in terms of its Hamiltonian and is always related to the high energy spectrum.  Any model is thought of as a relevant deformation of a fixed point model, which controls it's asymptotic spectrum.  Thus, in AdS/CFT the c.c. is part of the definition of the model and has nothing to do with quantum corrections to a classical approximation.  How can this be compatible with the effective bulk field theory idea that the c.c. can be renormalized and gets both low and high energy contributions?   Many years ago\cite{tom} one of the present authors speculated that models with AdS radius much larger than any fundamental scale, which had a spectrum of low lying operators compatible with a gap between operators dual to excitations of "bulk gravity" and the typical exponentially rising density of states of a CFT, were all relevant perturbation of superconformal fixed points.  The shorthand term for such models is "CFTs with a large radius dual".  Recently, Ooguri and Vafa\cite{ov} have made progress in proving this conjecture.  In the reinterpretation of \cite{tbov}, the meaning of the Ooguri-Vafa result is that one cannot construct large radius SUSY violating duals as the near horizon limit of stacks of branes in Minkowski space, including branes wrapped around cycles in the compact dimensions.  The essential point is that SUSY violating stacks of branes are classified by small discrete groups and cannot generate the large back-reaction necessary to creating a large radius near horizon limit with the geometry of AdS.

This is also consistent with previous arguments\cite{tom}.  One can construct new examples of large radius AdS/CFT dual pairs by "orbifolding" existing models\cite{klsv}.  Whenever the orbifold preserves some SUSY, the procedure works, while it fails if SUSY is violated\footnote{A possible exception is the orbifolding of the $AdS_7 \times S_4$ solutions of M-theory, where a bulk analysis by S. Thomas shows no instability.}.   The work of \cite{igor} has clarified how this occurs.  At weak 't Hooft coupling, the orbifold field theories have non-zero $\beta$ functions for multiple trace operators at leading order in the large $N$ expansion.  This corresponds to deviations from the orbifold at classical order in string theory.  It has been argued that renormalization group flow leads back to the supersymmetric parent theory.  The authors of\cite{aharony}  have shown how this mechanism works at large 't Hooft coupling.  The multiple trace operators are products of relevant single trace operators with special dimensions ($d/2$ for the double trace example) for which the conventional Dirichlet boundary conditions lead to logarithmic behavior near the boundary.  To eliminate dependence on the arbitrary scale in the logarithm, the boundary condition must be changed to one that implies that a multiple trace operator marginal operator has been added to the Lagrangian at leading order in $N$.

 Similarly, breaking SUSY by relevant operators does not,generically, produce new large radius conformal fixed points\cite{holorg}.  Again there's a class of possible exceptions \cite{bobev}.   However, these correspond to solutions of consistent truncations of ten or eleven dimensional SUGRA, and no one has checked the absence of tachyons violating the Breitenlohner-Freedman (BF) bound in the Kaluza-Klein spectrum.  Since the solutions always involve an internal manifold with size of order the AdS radius, there is no parametric lifting of the KK spectrum, which could provide an easy proof that the KK modes are tachyon free.  In addition to possible BF  violating tachyons in the higher KK modes, one must also search for the kind of instability encountered in the SUSY violating orbifold theories.
 
 There is a beautiful paper\cite{giombi}, which addresses all of these issues for a particular class of perturbations of special superconformal boundary theories in $2 +1$ dimensions. There is however one issue, discussed by these authors, which has not been settled. The SUSic parent theory has a moduli space, which is lifted by the perturbation, and one cannot tell whether the potential for that moduli space, remains bounded from below as one flows to the infrared, although it appears stable near the UV fixed point.  If it is not stable, this should be interpreted as a Big Crunch in the bulk\cite{hhharlowbarbonmalda} rather than a flow to a true CFT.  Presumably the instability (if it exists) is stabilized by a dangerous irrelevant operator from the IR point of view, as proposed in \cite{hh}.  The resulting theory at the stable vacuum is not conformal invariant.  Of course, the question of whether this moduli space instability exists is still open, so this remains a promising class of exceptions to what is now known as the AdS swampland conjecture.

Finally, extensive searches of models with large central charge in $1 + 1$ dimensions, have failed to reveal any examples of large radius models\cite{???}, as have searches through large $N$ models with large representations of the gauge group.  We want to emphasize that finding an ironclad example of a large radius non-SUSic CFT, would not affect our basic argument at all, but would instead produce a puzzle.  It would still be true that the c.c. was a parameter in the theory, characterizing its high energy spectrum, rather than a quantity that received corrections from calculations of bulk energy density.  However, in bulk effective field theory, it is extremely probable that such bulk corrections to the effective c.c. would exist and have UV divergences. We should note that the models of\cite{giombi} evade this argument because they violate SUSY only by boundary conditions in AdS.  Thus, bulk contributions to the c.c. still vanish exactly in these models.  They are reminiscent of the two dimensional models of \cite{silverfreedman}, which also involve double trace perturbations.  

If we accept the conjecture that all large radius CFTs arise from SUSY broken only at the boundary of AdS space by multiple trace operators, then a rather satisfactory agreement between bulk and boundary arguments emerges.  The small c.c. in the bulk cannot be renormalized because of SUSY, and in the boundary theory it is just a parameter characterizing the number of fundamental degrees of freedom, which is obviously an input to the theory and cannot be "corrected". 

We therefore think of ideas like\cite{arkaniibanez}, which attempt to calculate "finite low energy corrections to the c.c." as {\it fundamentally incompatible} with the lesson that AdS/CFT teaches us about the nature of the c.c. .  It does not, in our opinion, make sense to combine them with the Ooguri-Vafa conjecture, to obtain bounds on parameters in the standard model.

\section{What's the Matter With Quantum Field Theory?}

In this section, we want to collect a number of arguments, mostly coming from classical GR, which indicate precisely how quantum field theory fails to account for the properties of the boundaries of causal diamonds, and give us hints about how field theory emerges from a valid model of quantum gravity.

The first classical fact for which field theory fails to give an explanation, is a direct consequence of the negative specific heat of black holes in Minkowski or dS space, and small black holes in AdS space.  Let us drop a particle of small mass $m$ into a very large black hole\footnote{In AdS space, we are always talking about the large radius limit and about very large black holes, which are nonetheless much smaller than the AdS radius.}.  The increase in entropy caused by this event is proportional to $(M + m)^{\frac{d - 2}{d - 3}} - 
M^{\frac{d - 2}{d - 3}} \approx \frac{d - 2}{d - 3} m M^\frac{1}{d - 3}$.  This is enormous.  The question one must ask is the origin of this enormous increase of entropy.  There is probably universal agreement that one must answer the question in terms of a subset of degrees of freedom localized near the horizon, whose Hilbert space describes both the infalling particle and the black hole.  The increase in entropy implies that the initial state must not be a typical state in the Hilbert space, and so must satisfy a large number of linear constraints.  The process of infall and thermalization must correspond to the removal of these constraints by exciting frozen degrees of freedom.

Since the work of 't Hooft \cite{brickwall} field theorists have thought of black hole entropy in terms of very short wavelength degrees of freedom living on a stretched horizon, a time-like hyperbola that hugs the horizon.  These excitations of quantum fields in a black hole background have very low asymptotic energy, despite their shortwavelength.  On the other hand, in the coordinate system of an infalling object, they have very high energy.  Indeed, if we compute the renormalized stress tensor of the quantum fields near the horizon, we get very large values {\it unless there is maximal entanglement between short wavelength degrees of freedom on the two sides of the horizon}.   Thus, if the process of infall excites the state of these near horizon DOF, there will be a {\it firewall} experienced by infalling objects.   On the other hand, we can make exactly the same statements about a Rindler horizon in Minkowski space, which is geometrically similar to the Schwarzschild horizon for large mass.  In field theory, a low energy particle falling through the Rindler horizon does absolutely nothing to the entanglement of the shortwavelength DOF on either side of the horizon.  We believe that this is a strong argument that the explanation of the increase in entropy in terms of field theory Hilbert spaces is incorrect, and that there is NO explanation of this increase in the field theory language.

A complementary problem emerges when we study the entropy of black holes in dS space.
Here, the issue is a large {\it decrease} in entropy.   In Planck units, the radii of the two horizons of the dS Schwarzschild black hole satisfy
\begin{equation} R^2 - R_+^2 - R_-^2 = R_+ R_- \approx 2RM, \end{equation} when the mass is much less than the maximal Nariai mass.  $R$ is the empty dS radius.  The Gibbons-Hawking entropy thus decreases by an amount $2\pi RM$, which is exactly what we expect for the thermal probability of finding the black hole in the dS thermal ensemble.  In other words {\it a localized object in dS space leads to a decrease in entropy}.  Thus it must be a constrained state of the system.  Note that this is true even if the object is not a black hole, since it still produces a Schwarzschild deformation of the gravitional field.  In that case there would be no local increase in entropy from the black hole horizon.  We have a derivation of the temperature of dS space that does not depend on quantum field theory.  

Field theory has no way of accounting for this decrease.  Again, many have tried to interpret the entropy of dS space as entanglement entropy with the region outside the cosmological horizon.  However, there is a lot of evidence that the entanglement explanation of the entropy of horizons is just a consequence of using the thermofield double to purify the thermal state of the horizon\cite{israelmaldadks} .  One can purify the state of any system by entangling it with any other system whose Hilbert space is as least as large.  The global dS geometry is precisely the analog of the Kruskal extension of the Schwarzschild black hole.  It's a mathematical trick for calculating properties of a thermal system.  Field theory in dS space would tell us that the entropy of a localized system should simply be added to the mysterious entropy of the dS horizon.  If one tried to interpret that entropy in terms of entangled states of field modes localized near the cosmological horizon, there would be no change in that entropy when we introduced a state with field energy localized near the origin.   The field theory picture is even more confusing when we introduce the localized energy at a point somewhat displaced from the origin.  The field theory calculation says that the additional energy introduced some entropy, increasing the total, but that entropic subsystem disappears in a time of order $R$ into the horizon, leading to a decrease in the total entropy.  Recall that when the field theory is formulated in static coordinates, the disappearing system never actually exits the causal patch, so this process violates the second law of thermodynamics.   The only excuse for this is that in the formal limit $L_P \rightarrow 0$, the horizon entropy is infinite, but there's no account of the huge finite increase that occurs when the object merges with the horizon.  In other words, the small gravitational backreaction of the localized object on the global geometry, is not accounted for by any entropy count in field theory.  

Another disturbing feature of the entropy of empty dS space, is that it doesn't go away in the limit $R\rightarrow\infty$.  Indeed, it becomes infinite.  We are used to treating quantum gravity in Minkowski space in terms of an S-matrix and among the assumptions of S-matrix theory is that the Fock vacuum of asymptotic particles is the unique zero energy state.  In four dimensions we know that this assumption is wrong.  Infrared divergences, whose inclusive effect has been understood since the work of Weinberg in the 1960s\cite{weinir} imply that all matrix elements of the gravitational S matrix in Fock space, vanish.  Weinberg showed that inclusive cross sections with a cut on the missing soft graviton energy were insensitive to this problem and this certainly resolves the "practical" issue of comparing the predictions of the theory to experiment.  Nonetheless, anyone who is concerned about unitarity of the S operator for black hole production and decay, must first decide on the nature of the Hilbert space in which it operates.  

String theorists have avoided thinking about this problem because the perturbative S matrix has finite matrix elements in Fock space, once one goes to sufficiently high dimension.  However, as the four dimensional case shows, the real issue has to do with infinite numbers of arbitrarily soft gravitons.  This is related then to the behavior of the perturbation series for very high orders, and we know that it diverges badly\cite{shenker}.  Indeed, when one thinks about the physics this S-matrix is supposed to describe, it becomes obvious that no perturbative treatment of this question is adequate.  For example, it is widely believed\cite{bfetal} that scattering of two gravitons, at an impact parameter smaller than the Schwarzschild radius of the center of mass energy, will produce a black hole.  The gravitational S-matrix in any number of dimensions, thus describes processes in which scattering of a finite number of particles produces a collection of large black holes, which can orbit around each other emitting gravitational bremstrahlung, coalesce, and ultimately decay.  Can anyone seriously claim that the finiteness of the perturbative S matrix elements
of a badly divergent perturbation series settles the question of whether the soft graviton state produced in this process is a normalizable state in Fock space?  The only non-perturbative model of gravitational scattering in Minkowski space, of which we are aware\footnote{In \cite{HSTAdS} we argued that one cannot address this question by taking limits of CFT correlators.  We will comment on this below.} is the large $N$ limit of Matrix Theory\cite{bfss}.  In this model, soft gravitons correspond to very small matrix blocks, which carry very small transverse momentum.   As $N\rightarrow\infty$ it's clear that we must examine the question of whether the unitary scattering matrix of the finite $N$ theory decouples from the states with infinite numbers of such small blocks.  There is absolutely no indication that it will do so.

It is thus plausible to conjecture that the limit of the high entropy vacuum ensemble of dS space is the correct description of the soft graviton part of the space of scattering states.
In four dimensions, Fadeev and Kulish proposed a method for computing IR finite amplitudes for quantum gravity, and there has been much recent work on this\cite{fketal}.
This prescription appears to describe an IR finite perturbation theory, but does not deal with the divergence of the perturbation series itself.

%If it is true that Minkowski models of quantum gravity only exist when there is exact %SUSY, then it is likely that there are models of dS space only in four dimensions.  %Many four dimensional SUGRA models with chiral multiplets have dS solutions.  No %SUGRA models in higher dimensions have any\cite{juannogoetal}.  

We will now argue that in approaching Minkowski space from the $AdS_d$ side, one is led to a similar picture of an infinite entropy vacuum ensemble rather than a unique vacuum state.  As emphasized by Susskind and Polchinski\cite{susspol}, in order to study this limit one must concentrate on a single causal diamond, "the arena",  whose area is much smaller than $R^{d - 2}$ in $d$ dimensional Planck units.  We will make the assumption that the Hilbert space necessary to describe events in the arena is finite dimensional.  The proper time between the tips of the arena is $\ll R$. Consider a generic state obtained by acting on the CFT vacuum with a finite number of local operators at some global time prior to the past tip of the arena, but by an amount $\ll R$.  In the large $N$ limit we can describe the evolution of this state by Feynman-Witten diagrams, and we know that for most configurations of the boundary operators the lines of those diagrams have very low probability of entering the arena.   Thus, there are many states of the CFT, for a wide range of CFT energy, which must be the vacuum state when restricted to the arena.  This already shows us that we cannot identify the CFT Hamiltonian eigenvalue with the emergent Minkowski energy.  Many states, with different CFT energies, all have Minkowski energy which is approximately $0$.  

In particular, consider a global coordinate system whose $r = 0$ trajectory goes through the past and future tips of the arena.  These coordinates correspond, in the CFT to a particular choice of generator $K_0 + P_0$ in the conformal group.  Now consider a generator obtained by conjugating this with a non-compact generator of the conformal group, with very large "boost" parameter.  This moves the coordinates to a system centered at a point ${\bf r}$ very far from $r = 0$.  Consider a black hole of Schwarzschild radius much less than the distance from the origin to ${\bf r}$, traveling along the time-like geodesic centered at ${\bf r}$.  This black hole also looks like the vacuum in the arena.   On the other hand, this black hole is a typical microstate of the ensemble of states of the CFT, subject only to the constraint that the expectation value of the boosted Hamiltonian is fixed.  Thus, by Page's theorem, the probability that the density matrix of the arena is not maximally uncertain, is extremely small.   We conclude that, from the point of view of AdS/CFT, the Minkowski vacuum state of the arena is a maximally uncertain density matrix. 

We've thus reached the same conclusion by two very different routes, and by arguments that are not restricted to four dimensions.  The vacuum of Minkowski space, like that of dS space is a maximally uncertain density matrix, and localized objects in either of those space-times are perforce states satisfying a large number of constraints. 

\section{Conclusions}

We have argued that bulk quantum field theory implies results in apparent contradiction with classical gravitational reasoning, once one accepts the Bekenstein-Gibbons-Hawking formulae for the entropies of black hole and cosmological horizons.  How then does quantum field theory emerge from a theory of quantum gravity obeying these entropy laws? We believe that the clue to answering this question lies in the idea that localized states in a causal diamond are constrained states of a set of degrees of freedom living on the boundary of the diamond.  A useful cartoon of what we believe is the correct description is to imagine that the only degrees of freedom in the universe correspond to "very soft gravitons" penetrating a particle detector living on the holographic screens of causal diamonds.  The detector on a particular diamond covers the entire holographic screen, and in a generic state of the system, every pixel of the detector is lit up.  Now consider a set of constrained states in which pixels are turned off in a finite collection of annuli each of which is contained between two $d - 3$ cycles, both of which wind around the same point.

Now consider a smaller diamond, defined by a smaller interval of proper time along the geodesic between the tips of the original diamond.  This diamond will be assigned fewer pixels, since its holoscreen has smaller area.  In the simplest situation, all of the constrained annuli on the original diamond, will have images on the holoscreen of the small diamond.

\begin{figure}[h!]
\begin{center}
\includegraphics[width=12cm]{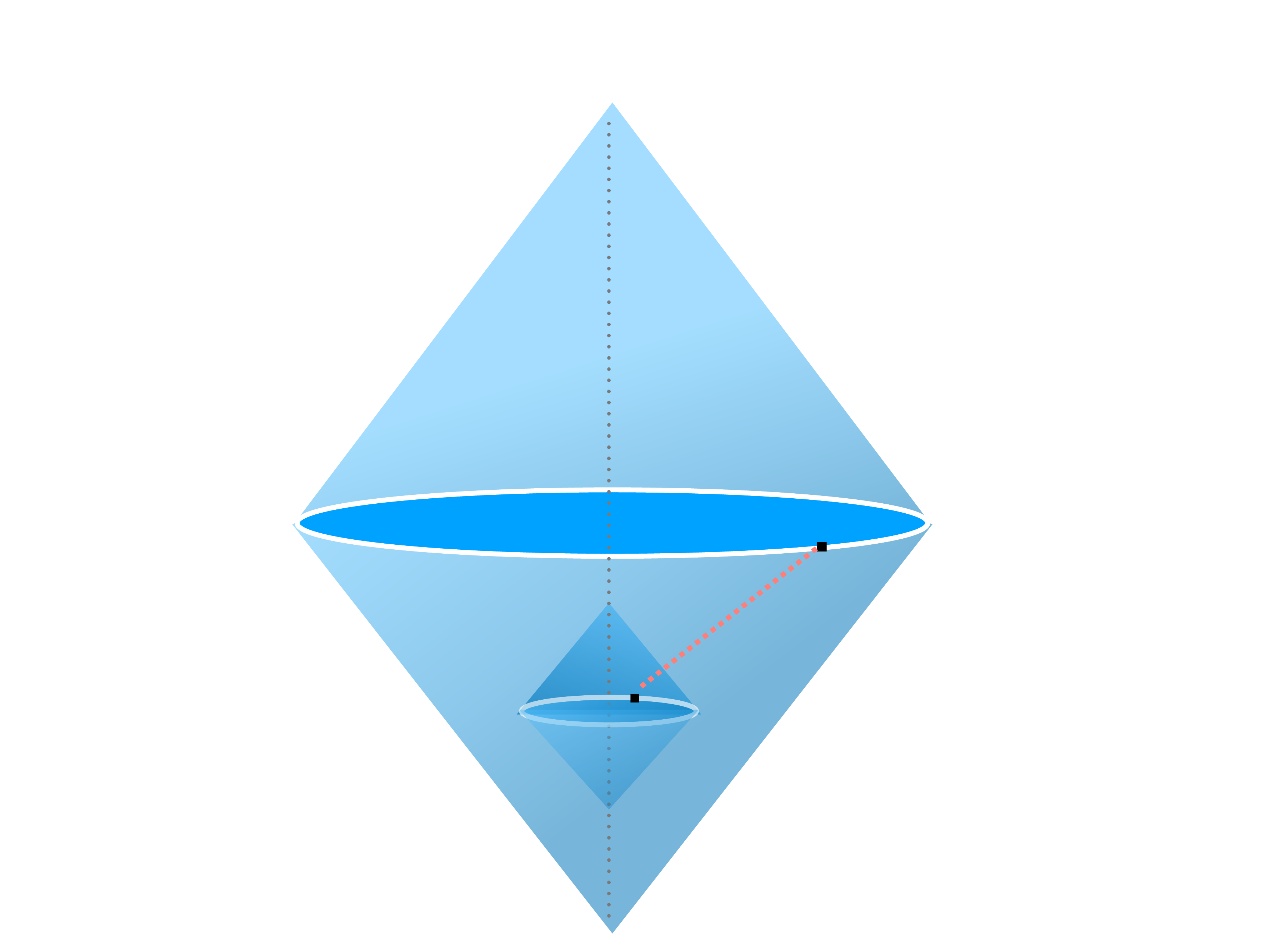}
\end{center}

\caption{ 
constraints are depicted as unlit pixels on the holoscreens}

\label{fig1}
\end{figure}

As shown in Figure 1, following the constraints between diamonds defines trajectories in the bulk space-time.  We consider these to be trajectories of jets of localized particles, by which we mean flows of energy and other quantum numbers.  The energies of the jets will be related to the number of unlit pixels.  Thus, if degrees of freedom live on holoscreens and localized objects are constrained states of those DOF, then one sees that they are localized by following the constraints in the bulk.  In a more general situation, some of the constraints on the original variables might be completely lost in the smaller diamond.  This would correspond to jets of particles that did not enter into that diamond.  One can begin to see the emergence of field theoretic Feynman diagrams from such a formalism.  The degrees of freedom missed by field theory are all of those pixel variables that lie outside of all the annuli. 

It is conceivable that one can interpret the boundary variables as field configurations that are pure gauge in the bulk and non-vanishing on the boundary.  There are many problems with such an approach.  Any field theoretic quantization of gravity fixes the gauge and so has propagating degrees of freedom in the bulk.  It's completely unclear how the field theory dynamics could be arranged so that physical degrees of freedom live only on causal diamond boundaries.  Indeed, the degrees of freedom on a smaller diamond are naturally thought of as a subset of those on a larger diamond, and so are localized in the smaller region only for a limited time.  It's completely unclear how to arrange such a situation for a field theory, which has degrees of freedom rigidly attached to points in space.
\begin{figure}[h!]
\begin{center}
\includegraphics[width=8cm]{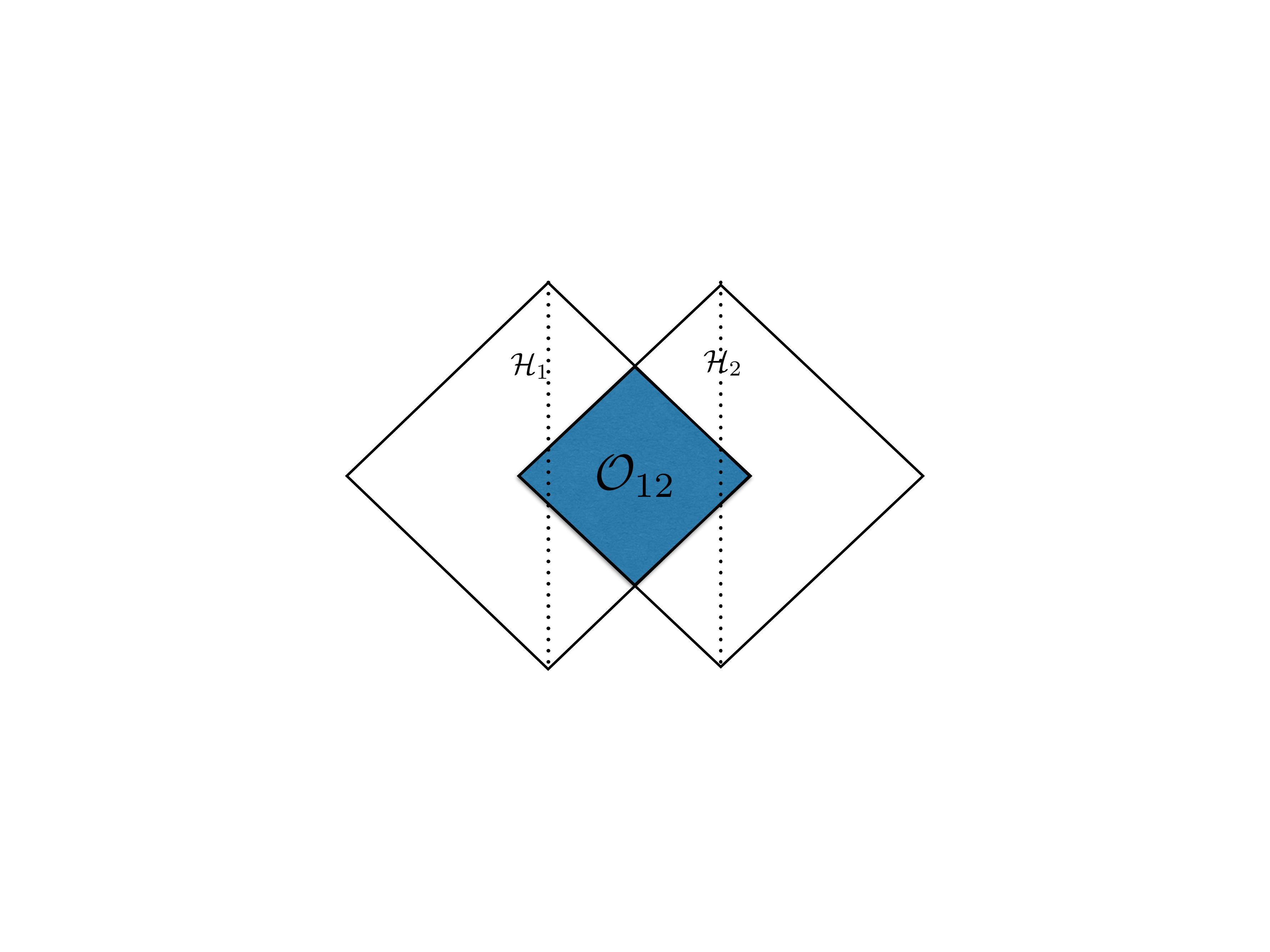}
\end{center}

\caption{ 
Intersecting diamonds}

\label{fig2}
\end{figure}

A dynamics that naturally incorporates a picture of degrees of freedom that move from holoscreen to holoscreen is proper time dynamics along a particular time-like trajectory, using time slices that remain inside each diamond for the proper time corresponding to that diamond.  The Hamiltonian must be time dependent (as appropriate for such causal slices) and couple together more degrees of freedom as time goes on.   Relativity can be incorporated by requiring equal entanglement spectra for the density matrices of overlap regions (Fig. 2) predicted by the dynamics along different trajectories.

  We are of course outlining the formalism of Holographic Space Time\cite{hst}.   That formalism is incomplete because the solution of the entanglement constraints for trajectories in relative motion has not been found.  The purpose of the present note is to point out a number of well agreed upon facts, where the field theoretic discussion falls short, but for which HST gives a straightforward explanation.

\vskip.3in
\begin{center}
{\bf Acknowledgments }\\
The work of T.Banks is {\bf\it NOT} supported by the Department of Energy, the National Science Foundation, the Simons or Templeton Foundations or FQXi. The work of W.Fischler is supported by the National Science Foundation under Grant Number PHY-1620610. We thank H.Nicolai, T.Padmanabhan, I Klebanov and N.Warner for pointing out extremely relevant earlier work, and helping us to correct errors in the original version of this manuscript.
\end{center}


\begin{thebibliography}{99}
\bibitem{weinbergcdl} S.~Weinberg,
``The Cosmological Constant Problem'',
Rev. Mod. Phys. {\bf61}, 1 and references therein
doi: 10.1103/RevModPhys.61.1;
S.~R.~Coleman and F.~De Luccia,
``Gravitational Effects on and of Vacuum Decay'',
Phys. Rev. {\bf D21}, 3305.
doi: 10.1103/PhysRevD.21.3305.
\bibitem{tp3}T.~Padmanabhan and Hamsa Padmanabhan, 
``Cosmic Information, the Cosmological Constant and the Amplitude of primordial perturbations'', Phys. Lett. {\bf B773}, 81-85 (2017) [arXiv:1703.06144]
 \bibitem{ted} T.~Jacobson,
 ``Thermodynamics of space-time: The Einstein equation of state,''
 Phys.\ Rev.\ Lett.\  {\bf 75}, 1260 (1995).
 \bibitem{BHJFSB}  J.~D.~Bekenstein,
  ``Black holes and entropy,''
  Phys.\ Rev.\ D {\bf 7}, 2333 (1973);
  %%CITATION = PHRVA,D7,2333;%%
  S.~W.~Hawking,
  ``Particle Creation by Black Holes,''
  Commun.\ Math.\ Phys.\  {\bf 43}, 199 (1975)
  [Commun.\ Math.\ Phys.\  {\bf 46}, 206 (1976)];
  %%CITATION = CMPHA,43,199;%%
    W.~Fischler and L.~Susskind,
  ``Holography and cosmology,''
  hep-th/9806039;
  %%CITATION = HEP-TH/9806039;%%
  R.~Bousso,
  ``A Covariant entropy conjecture,''
  JHEP {\bf 9907}, 004 (1999)
  [hep-th/9905177].
  %%CITATION = HEP-TH/9905177;%%
  \bibitem{bousso} R.~Bousso,
  ``Positive vacuum energy and the N bound'',
  JHEP 1{\bf1, 038},
  [hep-th/0010252]
  \bibitem{tp1} T.~Padmanabhan, 
  ``General relativity from a thermodynamic perspective'', Gen.Rel.Grav, {\bf46}, 1673 (2014) [arXiv:1312.3253]; see sec 6.
  \bibitem{tp2} T. ~Padmanabhan, 
  ``Distribution function of the Atoms of Spacetime and the Nature of Gravity'', Entropy {\bf17}, 7420-7452 (2015) [arXiv:1508.06286]; see sec 3
  \bibitem{hh}J.~ B.~Hartle and S.~W.~Hawking,
  ``Path Integral Derivation of Black Hole Radiance'',
  Phys. Rev. {\bf D13}, 2188; 
S.~M.~ Christensen and M~.J.~ Duff,
``Flat Space As A Gravitational Instanton''
Nucl.Phys. {\bf B146} (1978) 11-19; 
  See also sec 3.2 in T.~Padmanabhan, 
  ``Gravity and the Thermodynamics of Horizons'', Phys. Reports, {\bf406}, 49 (2005) [gr-qc/0311036]
  \bibitem{tom}  T.~Banks,
  ``Strings in a Landscape,'' Prepared for NATO Advanced Study Institute and EC Summer Sch Conference, p 3-7;
  %%CITATION = INSPIRE-673172;%%

T.~Banks,
  ``Landskepticism or why effective potentials don't count string models,''
  hep-th/0412129;
  %%CITATION = HEP-TH/0412129;%% 

T.~Banks and K.~van den Broek,
  ``Massive IIA flux compactifications and U-dualities,''
  JHEP {\bf 0703}, 068 (2007)
  doi:10.1088/1126-6708/2007/03/068
  [hep-th/0611185];
  %%CITATION = doi:10.1088/1126-6708/2007/03/068;%% 

T.~Banks,
  ``The Top $10^{500}$ Reasons Not to Believe in the Landscape,''
  arXiv:1208.5715 [hep-th].
  %%CITATION = ARXIV:1208.5715;%%,
  
  \bibitem{ov}H.~Ooguri and C.~Vafa,
  ``Non-supersymmetric AdS and the Swampland''
  Adv. Theor. Math. Phys. 21, 1787,
  arXiv:1610.01533 [hep-th]
 \bibitem{tbov} T.~Banks,
  ``Note on a Paper by Ooguri and Vafa,''
  arXiv:1611.08953 [hep-th].
  %%CITATION = ARXIV:1611.08953;%%,
 
   \bibitem{klsv}S.~Kachru and E.~Silverstein,
  ``4-D conformal theories and strings on orbifolds,''
  Phys.\ Rev.\ Lett.\  {\bf 80}, 4855 (1998)
  doi:10.1103/PhysRevLett.80.4855
  [hep-th/9802183];
  %%CITATION = doi:10.1103/PhysRevLett.80.4855;%%
  
  A.~E.~Lawrence, N.~Nekrasov and C.~Vafa,
  ``On conformal field theories in four-dimensions,''
  Nucl.\ Phys.\ B {\bf 533}, 199 (1998)
  doi:10.1016/S0550-3213(98)00495-7
  [hep-th/9803015].
  %%CITATION = doi:10.1016/S0550-3213(98)00495-7;%%
  \bibitem{igor}  A.~Dymarsky, I.~R.~Klebanov and R.~Roiban,
  ``Perturbative gauge theory and closed string tachyons,''
  JHEP {\bf 0511}, 038 (2005)
  doi:10.1088/1126-6708/2005/11/038
  [hep-th/0509132];
  %%CITATION = doi:10.1088/1126-6708/2005/11/038;%%
   A.~Dymarsky, I.~R.~Klebanov and R.~Roiban,
  ``Beta functions for double-trace couplings in orbifold gauge theories,''
  Int.\ J.\ Mod.\ Phys.\ A {\bf 20}, 6278 (2005);
  doi:10.1142/S0217751X05029307
  %%CITATION = doi:10.1142/S0217751X05029307;%%
   E.~Pomoni and L.~Rastelli,
  ``Large N Field Theory and AdS Tachyons,''
  JHEP {\bf 0904}, 020 (2009)
  doi:10.1088/1126-6708/2009/04/020
  [arXiv:0805.2261 [hep-th]].
  %%CITATION = doi:10.1088/1126-6708/2009/04/020;%%
  \bibitem{aharony} O.~Aharony, G.~Gur-Ari and N.~Klinghoffer,
  ``The Holographic Dictionary for Beta Functions of Multi-trace Coupling Constants,''
  JHEP {\bf 1505}, 031 (2015)
  doi:10.1007/JHEP05(2015)031
  [arXiv:1501.06664 [hep-th]].
  %%CITATION = doi:10.1007/JHEP05(2015)031;%%
  \bibitem{holorg} K.~Skenderis,
  ``Lecture notes on holographic renormalization,''
  Class.\ Quant.\ Grav.\  {\bf 19}, 5849 (2002)
  doi:10.1088/0264-9381/19/22/306
  [hep-th/0209067], and references cited therein.
  %%CITATION = doi:10.1088/0264-9381/19/22/306;%%
  
  \bibitem{bobev} T.~Fischbacher, K.~ Pilch and N.~P.~Warner,
    ``New Supersymmetric and Stable, Non-Supersymmetric Phases in Supergravity and Holographic Field Theory''                    
  [arXiv: 1010.4910 [hep-th]];
  N.~Bobev, A.~Kundu, K.~Pilch, Krzysztof and N.~P.~Warner, 
  ``Minimal Holographic Superconductors from Maximal Supergravity''
  JHEP {\bf03}, 064
  [arXiv: 1110.3454 [hep-th]];
  H.~Godazgar, M.~ Godazgar, O.~ Kruger, H.~Nicolai and K.~Pilch, 
  ``An SO(3)$\times$SO(3) invariant solution of $D=11$ supergravity''
  JHEP {\bf01}, 056
   [arXiv: 1410.5090 [hep-th]];
   T.~Fischbacher, H.~Nicolai and H.~Samtleben,
  ``Vacua of maximal gauged D = 3 supergravities,''
  Class.\ Quant.\ Grav.\  {\bf 19}, 5297 (2002)
  doi:10.1088/0264-9381/19/21/302
  [hep-th/0207206];
  %%CITATION = doi:10.1088/0264-9381/19/21/302;%%
  H.~Nicolai and H.~Samtleben,
  ``Chern-Simons versus Yang-Mills gaugings in three-dimensions,''
  Nucl.\ Phys.\ B {\bf 668}, 167 (2003)
  doi:10.1016/S0550-3213(03)00569-8
  [hep-th/0303213];
  %%CITATION = doi:10.1016/S0550-3213(03)00569-8;%%
  H.~Nicolai and H.~Samtleben,
  ``Kaluza-Klein supergravity on AdS(3) x S**3,''
  JHEP {\bf 0309}, 036 (2003)
  doi:10.1088/1126-6708/2003/09/036
  [hep-th/0306202].
  %%CITATION = doi:10.1088/1126-6708/2003/09/036;%%
  \bibitem{giombi}  S.~Giombi and E.~Perlmutter,
  ``Double-Trace Flows and the Swampland,''
  JHEP {\bf 1803}, 026 (2018)
  doi:10.1007/JHEP03(2018)026
  [arXiv:1709.09159 [hep-th]].
  %%CITATION = doi:10.1007/JHEP03(2018)026;%% S.~Giombi and E.~Perlmutter,
    \bibitem{hhharlowbarbonmalda}
  T.~Hertog and G.~T.~Horowitz,
  ``Towards a big crunch dual,''
  JHEP {\bf 0407}, 073 (2004)
  doi:10.1088/1126-6708/2004/07/073
  [hep-th/0406134];
  %%CITATION = doi:10.1088/1126-6708/2004/07/073;%%
  J.~L.~F.~Barbon and E.~Rabinovici,
  ``Holography of AdS vacuum bubbles,''
  JHEP {\bf 1004}, 123 (2010)
  doi:10.1007/JHEP04(2010)123
  [arXiv:1003.4966 [hep-th]];
  %%CITATION = doi:10.1007/JHEP04(2010)123;%%
  J.~L.~F.~Barbon and E.~Rabinovici,
  ``AdS Crunches, CFT Falls And Cosmological Complementarity,''
  JHEP {\bf 1104}, 044 (2011)
  doi:10.1007/JHEP04(2011)044
  [arXiv:1102.3015 [hep-th]];
  %%CITATION = doi:10.1007/JHEP04(2011)044;%%
   D.~Harlow,
  ``Metastability in Anti de Sitter Space,''
  arXiv:1003.5909 [hep-th];
  %%CITATION = ARXIV:1003.5909;%%
  D.~Harlow and L.~Susskind,
  ``Crunches, Hats, and a Conjecture,''
  arXiv:1012.5302 [hep-th];
  %%CITATION = ARXIV:1012.5302;%%
J.~Maldacena,
  ``Vacuum decay into Anti de Sitter space,''
  arXiv:1012.0274 [hep-th].
  %%CITATION = ARXIV:1012.0274;%%
  \bibitem{hh} T.~Hertog and G.~T.~Horowitz,
  ``Towards a big crunch dual,''
  JHEP {\bf 0407}, 073 (2004)
  doi:10.1088/1126-6708/2004/07/073
  [hep-th/0406134].
  %%CITATION = doi:10.1088/1126-6708/2004/07/073;%%
  \bibitem{silverfreedman} O.~Aharony, M.~Berkooz and E.~Silverstein,
  ``Nonlocal string theories on AdS(3) x S**3 and stable nonsupersymmetric backgrounds,''
  Phys.\ Rev.\ D {\bf 65}, 106007 (2002)
  doi:10.1103/PhysRevD.65.106007
  [hep-th/0112178];
  %%CITATION = doi:10.1103/PhysRevD.65.106007;%% 
  X.~Dong, D.~Z.~Freedman and Y.~Zhao,
  ``Explicitly Broken Supersymmetry with Exactly Massless Moduli,''
  JHEP {\bf 1606}, 090 (2016)
  doi:10.1007/JHEP06(2016)090
  [arXiv:1410.2257 [hep-th]].
  %%CITATION = doi:10.1007/JHEP06(2016)090;%%
  \bibitem{???} A.~Belin, C.~A.~Keller and A.~Maloney,
  ``Permutation Orbifolds in the large N Limit,''
  Annales Henri Poincare, 1 (2016)
  doi:10.1007/s00023-016-0529-y
  [arXiv:1509.01256 [hep-th]];
  %%CITATION = doi:10.1007/s00023-016-0529-y;%% 
   T. Banks unpublished.
  
  \bibitem{arkaniibanez}N.~Arkani-Hamed, S.~Dubovsky, A.~Nicolis and G.~Villadoro,
  ``Quantum Horizons of the Standard Model Landscape,''
  JHEP {\bf 0706}, 078 (2007)
  doi:10.1088/1126-6708/2007/06/078
  [hep-th/0703067 [HEP-TH]];
  %%CITATION = doi:10.1088/1126-6708/2007/06/078;%%
  
  L.~E.~Ibanez, V.~Martin-Lozano and I.~Valenzuela,
  ``Constraining Neutrino Masses, the Cosmological Constant and BSM Physics from the Weak Gravity Conjecture,''
  JHEP {\bf 1711}, 066 (2017)
  doi:10.1007/JHEP11(2017)066
  [arXiv:1706.05392 [hep-th]];
  %%CITATION = doi:10.1007/JHEP11(2017)066;%%

 L.~E.~Ibanez, V.~Martin-Lozano and I.~Valenzuela,
  ``Constraining the EW Hierarchy from the Weak Gravity Conjecture,''
  arXiv:1707.05811 [hep-th];
  %%CITATION = ARXIV:1707.05811;%%

E.~Gonzalo, A.~Herraez and L.~E.~Ibanez,
  ``AdS-phobia, the WGC, the Standard Model and Supersymmetry,''
  JHEP {\bf 1806}, 051 (2018)
  doi:10.1007/JHEP06(2018)051
  [arXiv:1803.08455 [hep-th]];
  %%CITATION = doi:10.1007/JHEP06(2018)051;%%
  
  Y.~Hamada and G.~Shiu,
  ``Weak Gravity Conjecture, Multiple Point Principle and the Standard Model Landscape,''
  JHEP {\bf 1711}, 043 (2017)
  doi:10.1007/JHEP11(2017)043
  [arXiv:1707.06326 [hep-th]].
  %%CITATION = doi:10.1007/JHEP11(2017)043;%%
  
  \bibitem{brickwall}G.~'T Hooft,
  ``Horizons'',
  Subnucl. Ser. {\bf41}, 179.
  From quarks to black holes: Progress in understanding the logic of nature. Proceedings, International School of
  subnuclear physics, Erice, Italy, August 29-September 7, 2003.
  [gr-qc/0401027].
  \bibitem{israelmaldadks}W.~Israel, ``Thermo Field Dynamics Of Black Holes,'' Phys. Lett. {\bf A 57}, 107 (1976);
  L.~Dyson, M.~Kleban and L.~Susskind, ``Disturbing implications of a cosmological constant'', JHEP {\bf10}, 011
  [hep-th/0208013 [HEP-TH]];
  J.~Maldacena, ``Eternal black holes in anti-de Sitter'', JHEP {\bf 04}, 021,
  hep-th/0106112.
  \bibitem{weinir} S. ~Weinberg, 
  ``Infrared photons and gravitons,'' 
  Phys. Rev. {\bf140}, B516 (1965).
  
  \bibitem{shenker}S.~H.~Shenker,
  ``The Strength of nonperturbative effects in string theory'', in 
  Brezin, E. (ed.), Wadia, S.R. (ed.): The large N expansion in quantum field theory and statistical physics* 809-819.
  \bibitem{bfetal}T.~Banks and W.~Fischler, 
  ``A Model for high-energy scattering in quantum gravity,''
hep-th/9906038; 
D.~M.~Eardley and S.~B.~Giddings, Phys. Rev. {\bf D66}, 044011 [grqc/0201034];
S.~B.~Giddings and S.~D.~Thomas, Phys. Rev. {\bf D65}, 056010 [hepph/0106219].
  
  \bibitem{HSTAdS}T.~Banks and W.~Fischler,
  ``Holographic Space-time Models of Anti-deSitter Space-times''
  arXiv: 1607.03510 [hep-th]; 
  T.~Banks and W.~Fischler, 
  ``Soft Gravitons and the Flat Space Limit of Anti-deSitter Space'',
  arXiv: 1611.05906 [hep-th].
  \bibitem{bfss}T.~Banks, W.~Fischler, S.~H. Shenker and L.~Susskind,
  ``M theory as a matrix model: A Conjecture'',
  Phys. Rev. {\bf D55}, 5112.
[hep-th/9610043].
\bibitem{fketal} P.~P.~Kulish and L.~D.~Faddeev,
  ``Asymptotic conditions and infrared divergences in quantum electrodynamics,''
  Theor.\ Math.\ Phys.\  {\bf 4}, 745 (1970)
  [Teor.\ Mat.\ Fiz.\  {\bf 4}, 153 (1970)];
  doi:10.1007/BF01066485
  %%CITATION = doi:10.1007/BF01066485;%%P.~P.~Kulish,
  ``Asymptotical states of massive particles interacting with gravitational field,''
  Teor.\ Mat.\ Fiz.\  {\bf 6}, 28 (1971);
  doi:10.1007/BF01037574
  %%CITATION = doi:10.1007/BF01037574;%%
   P.P. Kulish "Infrared divergences of quantized
  gravitational field", Zapiski Nauchnykh Seminarov
  LOMI, vol.77, 106 - 123 (1978).
  \bibitem{juannogoetal}See for example: J.~M.~Maldacena and C.~Nunez,
  ``Supergravity description of field theories on curved manifolds and a no go theorem,''
  Int.\ J.\ Mod.\ Phys.\ A {\bf 16}, 822 (2001)
  doi:10.1142/S0217751X01003935, 10.1142/S0217751X01003937
  [hep-th/0007018].
  %%CITATION = doi:10.1142/S0217751X01003935, 10.1142/S0217751X01003937;%%  
    
   \bibitem{susspol} 
  J.~Polchinski,
  ``S matrices from AdS space-time,''
  hep-th/9901076.
  L.~Susskind,
  ``Holography in the flat space limit,''
  AIP Conf.\ Proc.\  {\bf 493}, 98 (1999)
  doi:10.1063/1.1301570
  [hep-th/9901079].
   \bibitem{hst} T.~Banks,
  ``Holographic Space-Time: The Takeaway,''
  arXiv:1109.2435 [hep-th];
  %%CITATION = ARXIV:1109.2435;%%
 T.~Banks,
  ``TASI Lectures on Holographic Space-Time, SUSY and Gravitational Effective Field Theory,''
  arXiv:1007.4001 [hep-th];
  %%CITATION = ARXIV:1007.4001;%%
  
  T.~Banks and W.~Fischler, ``Holographic Inflation Revised", arXiv: 1501.01686 [hep-th];
  
  T.~Banks and W.~Fischler,
  ``The holographic approach to cosmology,''
  hep-th/0412097;
  %%CITATION = HEP-TH/0412097
 
  T.~Banks and W.~Fischler,
  ``Holographic cosmology,''
  hep-th/0405200;
  %%CITATION = HEP-TH/0405200
  T.~Banks and W.~Fischler,
  ``Holographic cosmology 3.0,''
  Phys.\ Scripta T {\bf 117}, 56 (2005)
  [hep-th/0310288];
  %%CITATION = HEP-TH/0310288
 T.~Banks and W.~Fischler,
  ``An Holographic cosmology,''
  hep-th/0111142;
  %%CITATION = HEP-TH/0111142
  T.~Banks and W.~Fischler,
  ``M theory observables for cosmological space-times,''
  hep-th/0102077.

\end{thebibliography}
\end{document}